# Outcome-Driven Clustering of Acute Coronary Syndrome Patients using Multi-Task Neural Network with Attention

Eryu Xia[a], Xin Du[b], Jing Mei[a], Wen Sun[a], Suijun Tong[a], Zhiqing Kang[c], Jian Sheng[c], Jian Li[c], Changsheng Ma[b], Jianzeng Dong[b,d], Shaochun Li[a]

[a] *IBM Research – China, IBM, Beijing, China*
[b] *Department of Cardiology, Beijing Anzhen Hospital, Capital Medical University, National Clinical Research Center for Cardiovascular Diseases, Beijing, China*
[c] *AstraZeneca China*
[d] *Department of Cardiology, The First Affiliated Hospital, Zhengzhou University, Zhengzhou, Henan, China*

**Abstract**

*Cluster analysis aims at separating patients into phenotypically heterogenous groups and defining therapeutically homogeneous patient subclasses. It is an important approach in data-driven disease classification and subtyping. Acute coronary syndrome (ACS) is a syndrome due to sudden decrease of coronary artery blood flow, where disease classification would help to inform therapeutic strategies and provide prognostic insights. Here we conducted outcome-driven cluster analysis of ACS patients, which jointly considers treatment and patient outcome as indicators for patient state. Multi-task neural network with attention was used as a modeling framework, including learning of the patient state, cluster analysis, and feature importance profiling. Seven patient clusters were discovered. The clusters have different characteristics, as well as different risk profiles to the outcome of in-hospital major adverse cardiac events. The results demonstrate cluster analysis using outcome-driven multi-task neural network as promising for patient classification and subtyping.*

## Introduction

Precision medicine is a healthcare approach which aims at developing more effective ways to improve health and treat disease by taking individual traits into account [1]. One attempt toward precision medicine is to provide the best available care for patients based on their disease subtypes within a disease of common biological basis. Patient cluster analysis comprises a solid step towards precision medicine, which fulfils the task of disease classification and subtyping [2]. Cluster analysis has been used for subgroup analysis of type 2 diabetes [3], accurate phenotyping of heart failure and related syndromes [4,5], as well as identifying meaningful patient clusters for developing specific treatment programs in geriatric stroke patients [6]. The results support cluster analysis as a useful tool to discover disease classes and subtypes, which can inform therapeutic strategies like individualizing treatment regimens and providing prognosis insights.

Cluster analysis is performed based on a similarity or distance measure. Commonly used similarity measures include Euclidean distance, cosine similarity, Jaccard similarity, and so on. Traditionally, as no associating outcome measure is available for cluster analysis, the methods are unsupervised, and thus the similarity measure takes all patient characteristics as equally important, the results of which are less desired when we target the clustering results at reflecting specific patient traits. It has been recognized that patient similarities for cluster analysis are commonly context-based and are sometimes associated with clinical outcomes of interest. Outcome-driven clustering (sometimes referred to as 'semi-supervised clustering' or 'supervised clustering') is applied when outcome measures are available and can serve as a noisy surrogate for the (unobserved) target cluster [7], which has been proven useful in patient cluster analysis for precision cohort finding [8] and clinical decision support [9].

Neural network has been increasingly used as a successful data modeling paradigm, which solves tasks such as pattern recognition and classification through a learning process and has been recently used in medical informatics research for representation learning [10]. Multi-task learning is a strategy where multiple learning tasks are solved at the same time to benefit from their commonalities and contrasts. Neural networks adapt to multi-task learning intuitively by designing specific network structure and cost function [11]. Though useful, neural network is often criticized for lack of interpretability. Attention mechanism is thus introduced to neural network increase model interpretability as well as performance [12] and has been applied to healthcare research [13]. Thus, a joint use of the three techniques would facilitate representation learning leveraging information from different tasks in an interpretable manner.

Acute coronary syndrome (ACS) is a syndrome due to sudden decreased coronary artery blood flow. A treatment objective of ACS is to prevent major adverse cardiac events (MACE) during hospitalization. ACS can be classified into ST elevation myocardial infarction (STEMI), non-ST elevation myocardial infarction (NSTEMI), and unstable angina (UA) by cardiac marker and manifestation of ST-elevation in electrocardiogram. However, exploration of biomarkers for disease classification and subtyping has never stopped [14,15]. In this study, we presented outcome-driven clustering of ACS patients based on biomarkers as well as clinical indicators. We desired using patient state (which is an abstract characterization of patient traits regarding the disease) for clustering and decided on four outcome measures as surrogates to indicate the patient state: antiplatelet treatment, beta-blockers treatment, statins treatment, and in-hospital MACE. The four measures are supposed to reflect different facets of the patient state. Therefore, a joint consideration would enable a more comprehensive and targeted depiction of the patient state. Cluster analysis has been conducted on ACS patients to discover symptom clusters [16], assess the differences in mortality between symptom clusters [17], discover clusters of different lifestyle risk factors [18], and to detect critical patients using medical parameter time series [19]. However, all the above studies are unsupervised and none of them use neural network as the modeling framework.

In this study, we conducted outcome-driven patient clustering on hospitalized ACS patients, identified underlying patient clusters, and profiled the cluster characteristics, especially risk factors to in-hospital MACE. Novelty of our study includes: (1) using outcome-driven cluster analysis to guide cluster analysis; (2) using multi-task neural network to learn a multi-faceted representation of patient characteristics; and (3) attention mechanism was introduced to the neural network model to increase model interpretability and facilitate feature importance profiling.

## Methods

### Cohort construction

The multi-center retrospective cohort study was conducted at 38 urban and rural hospitals in China. Adult hospitalized patients (aged ⩾18 years) with a final diagnosis of ACS identified at the time of death or discharge were included. Each hospital enrolled the first five consecutive patients on a monthly basis from January 1, 2008 through December 31, 2015. We excluded patients who: (1) had potentially lethal diseases (e.g., incurable cancer, decompensated cirrhosis, multisystem organ failure); (2) had an expected life span below 12 months; or (3) died within 10 minutes of arrival at the hospital. A patient needs to have age, gender, ACS type recorded to be included in the analysis. A total of 26,986 patients were included in this study.

### Feature construction

Patient data was identified and reviewed by trained investigators to record clinical information. We included 41 patient characteristics as features, including: disease types (ACS type and Killip class), demographics, personal disease history, comorbidities, habits, laboratory test results, and procedures. Data outliers determined based on clinical knowledge were removed and represented as missing data. Missing values were imputed by multiple imputation utilizing the 'mice' package in R [20]: continuous variables by predictive mean matching, binary variables by logistic regression, and a proportional odds model for ordinal variables. We conducted one-hot encoding on categorical variables with more than two categories, and standardized continuous values by removing the mean and scaling to unit variance.

### Classification with multi-task neural network model

Multi-task neural network with attention was used as the framework. A schematic representation of the neural network design is shown in Figure 1. The attention layer was implemented as a hidden layer with softmax activation, with the same number of nodes as the shared input layer. Attention was added to the attention layer by multiplying element-wise with the shared input layer.

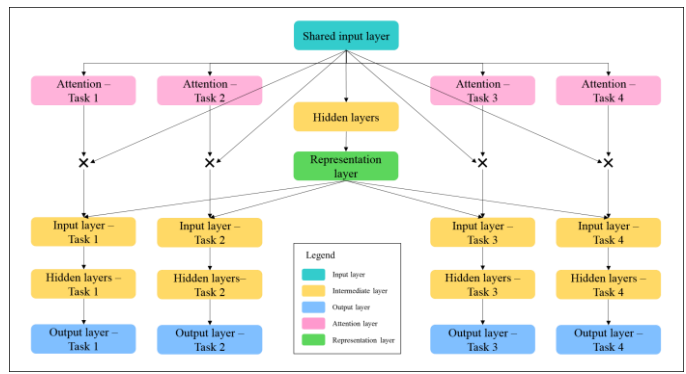

*Figure 1 - Illustration of the neural network design.*

Following the shared input layer are hidden layers, then a representation layer which learns a joint patient representation. The four classification tasks are optimized simultaneously. For each task, the input for the task was formed by concatenating the representation layer with the features in the shared input layer after adding attention to each feature. It then reaches the final output layer after adding hidden layers in between.

During training, features mentioned in the 'Feature construction' section were used as input for in the shared input layers, and the ground truth of the four patient traits (MACE, antiplatelet treatment, beta-blockers treatment, and statins treatment) were used as output in the four output layers for each task.

Binary cross entropy loss was used for each classification task. For task $j$ ($j$ in 1, 2, 3, and 4), the task weight is denoted as $w_j$, the ground truth and predicted probability for an instance $i$ are denoted as $y_{ij}$ and $\widehat{y_{ij}}$ respectively, and $H(y_{ij}, \widehat{y_{ij}})$ is the binary cross entropy loss. The cost function for neural network we used is defined as:

$$C \triangleq \frac{1}{n} \sum_{i=1}^{n} \sum_{j=1}^{4} w_j H(y_{ij}, \widehat{y_{ij}})$$

where n is the number of training samples. In our experiments, we assigned equal weight to all classification tasks.

Parameters were optimized using 'Adam'. Training was conducted with a batch size of 512 and 50 epochs. Class weights were added to balance the biased proportion of positive and negative cases respectively for all four tasks.

To validate the performance of the neural network for the classification tasks, cross validation was conducted 10 times by each randomly splitting data into training set and validation set at a ratio of 4:1. For patient clustering, all samples were used for neural network training.

### Post-classification analysis workflow

Analysis after classification with multi-task neural network model includes three steps: (1) evaluating neural network classification performance; (2) clustering patients using values from the representation layer; and (3) profiling risk factors for in-hospital MACE in each patient cluster using the attention values.

### Patient clustering

For each patient, values of the representation layer after training were used as the vector for clustering, which is a 32-dimension vector. K-means was used for patient clustering. Model selection was conducted using Bayesian Information Criteria to choose model from a range of different K (number of clusters) settings (2 to 15). We selected K = 7 for K-means clustering.

### Implementation

Cohort construction, feature construction and post-classification analysis were conducted using R 3.4.1. Neural network training and analysis were conducted using Python 2.7.14, Keras 2.2.4, and Theano 1.0.3.

# Results

## Neural network classification performance

We used the multi-task neural network for the four selected classification tasks. Proportions of positive cases in four classification tasks are shown in Table 1, which are largely imbalanced. Classification performances are evaluated by AUROC (area under the receiver operating characteristics) and AUPRC (area under the precision recall curve) on the validation set in a cross validation setting (Table 1). From the results, MACE and antiplatelet treatment were best classified while beta-blockers treatment has the lowest classification performance. The results suggest that the learned neural network model is a greater reflection of the patient states corresponding to MACE and antiplatelet treatment.

*Table 1 - Neural network classification performance*

| Task | Positive case | Performance | | |
|---|---|---|---|---|
| | | AUROC | AUPRC (class 0) | AUPRC (class 1) |
| MACE | 3.54% | 0.8602 (0.0141) | 0.9926 (0.0007) | 0.2924 (0.0551) |
| Antiplatelet treatment | 80.50% | 0.8634 (0.0078) | 0.5799 (0.0197) | 0.9640 (0.0038) |
| Beta-blockers treatment | 68.87% | 0.6881 (0.0131) | 0.5035 (0.0199) | 0.8184 (0.0097) |
| Statins treatment | 89.24% | 0.7725 (0.0167) | 0.2842 (0.0275) | 0.9635 (0.0045) |

Note: In performance cells, the numbers denote mean value (standard deviation).

## K-means clustering

K-means was conducted on the representation layer to cluster patients into seven groups. Clustering results are visualized in Figure 2, where t-SNE was conducted to reduce the 32-dimension data in the representation layer to 2 dimensions and used color to denote the assigned cluster membership. We see clear separation of the clusters in the low dimension representation.

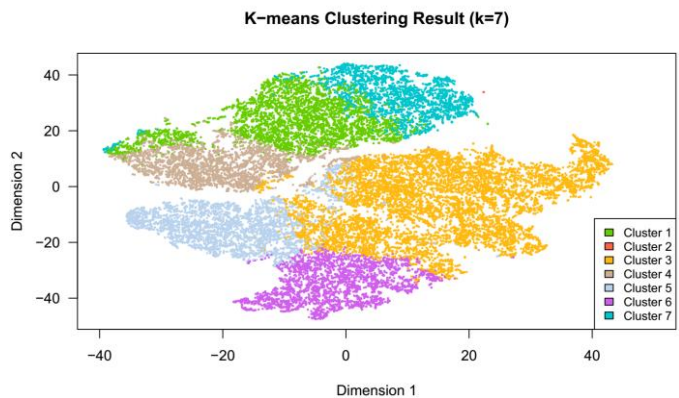

*Figure 2 - Visualization of patient clustering using t-SNE plot.*

Distribution of cluster size, and the properties relating to classification tasks are shown in Table 2. Clusters were organized based on descending MACE onset rate. The seven clusters have different cluster size, MACE rate and treatment rate. Cluster 2 has only 5 samples and is thus not included in later comparison. The largest cluster (9,889/26,986; 37%) had the highest MACE rate, and lower treatment rate compared to the overall cohort.

*Table 2 - Distribution of cluster size and classification labels*

| | Size | MACE | Anti-platelet treatment | Beta-blockers treatment | Statins treatment |
|---|---|---|---|---|---|
| Overall | 26,986 (100%) | 955 (3.5%) | 21,708 (80.5%) | 18,572 (68.9%) | 24,066 (89.2%) |
| Cluster 3 | 9,889 (37%) | 760 (7.7%) | 7,768 (78.6%) | 6,210 (62.8%) | 8,496 (85.9%) |
| Cluster 4 | 3,163 (12%) | 74 (2.3%) | 3,139 (99.2%) | 2,361 (74.6%) | 3,064 (96.9%) |
| Cluster 1 | 4,353 (16%) | 82 (1.9%) | 4,346 (99.8%) | 3,225 (74.1%) | 4,284 (98.4%) |
| Cluster 7 | 2,709 (10%) | 23 (0.8%) | 2,708 (100.0%) | 2,292 (84.6%) | 2,690 (99.3%) |
| Cluster 6 | 3,212 (12%) | 14 (0.4%) | 1,119 (34.8%) | 1,788 (55.7%) | 2,214 (68.9%) |
| Cluster 5 | 3,637 (13%) | 2 (0.1%) | 2,623 (72.1%) | 2,691 (74.0%) | 3,313 (91.1%) |
| Cluster 2 | 5 (0%) | 0 (0.0%) | 5 (100.0%) | 5 (100.0%) | 5 (100.0%) |

Note: In each cell, the numbers denote count (proportion). In column 'Size', the proportion is the proportion in the overall patient cohort. In other columns, the proportion is the proportion in the specified cluster.

Profiles of patient clusters were analyzed. Notable features of each patient cluster are presented in Table 3. Specifically, Cluster 3 has more severe conditions as is shown by the highest average age, proportion of patients with elevated cardiac enzyme levels, and the lowest average left ventricular ejection fraction (LVEF [%]). Cluster 4 also has comparatively severe condition as is shown by the Killip class. Cluster 1 does not show severe disease, but the MACE rate is still high, which is potentially associated with bad living habits (highest proportion of current smoker and current alcohol drinker) of this cluster. Cluster 7 is featured by the highest proportion of STEMI patients and lowest proportion of UA patients. Though STEMI patients are far more prone to in-hospital MACE compared to UA and NSTEMI, this cluster is not associated with a high MACE rate, potentially as a combined effect of the less complicated disease manifestation and the high level of treatment. Cluster 6 has the highest proportion of UA patients, and are less prone to MACE even though they have the lowest treatment rates. Cluster 5 is featured by the low disease severity, and correspondingly, the lowest MACE rate.

## Risk factors for MACE in each patient cluster

For each patient, a feature's attention value for the MACE classification task from the neural network is used as its importance in predicting MACE. For each patient cluster, a feature's importance is calculated as the average attention value of all patients in the cluster. Feature importance in each patient cluster is shown in Table 4. Features with different importance or high clinical relevance are selectively listed. The largest value in each row is shown in bold. Different clusters have different feature importance, indicating different risk profiles. As an example, smoking is a more important risk factor for MACE in Cluster 1 and Cluster 7, current comorbidity of hypertension more important for Cluster 4, 5 and 6, age and systolic blood pressure are more important in Cluster 3 than in other clusters.

*Table 3. Profile of patient clusters*

| Cluster | Has the highest | Has the lowest |
|---|---|---|
| 3 | MACE rate<br>Age<br>Proportion of NSTEMI patients<br>Cormobidity of atrial fibrillation<br>Elevated cardiac enzyme levels | LVEF (%) |
| 4 | Killip class<br>History of myocardial infarction | |
| 1 | Current smoker and current alcohol drinker | |
| 7 | Antiplatelet treatment<br>Beta-blockers treatment<br>Statins treatment<br>Proportion of STEMI patients | Proportion of UA patients<br>Killip class<br>Proportion of patients with cormobidity<br>Proportion of patients with disease history |
| 6 | Proportion of female patients<br>Proportion of UA patients | Antiplatelet treatment<br>Beta-blockers treatment<br>Statins treatment |
| 5 | LVEF (%)<br>Cormobidity of hypertension<br>History of vascular disease<br>History of established coronary artery disease<br>History of percutaneous coronary intervention<br>History of coronary artery bypass grafting<br>History of other conditions confirmed by computed tomography angiography | MACE rate<br>Elevated enzyme levels |

## Discussion

**Attention mechanism and feature importance.** Attention model in neural network is inspired by brain's neural mechanism of attention and is simplified here as: in each sample, including a numerical weight ('attention value') for each predictor associated with each outcome. When we normalize the weights for each sample to be all larger than 0 and have a sum of 1, the attention values look similar to a probability distribution to show the feature importance. In our study, we consider the outcome of in-hospital MACE. For each feature, we calculate the average attention value of patients in a cluster and use it as its importance in this cluster. When a feature has an importance larger than 0, we regard it as a risk factor to the outcome, where the feature importance is taken as the importance of the risk factor. The attention mechanism makes the neural network models, otherwise 'black boxes', interpretable to some degree, but is still less clear compared to logistic regression models, where both the feature importance and the action directionality (positive or negative impact) are shown with odds ratio, confidence intervals, and coefficient p-values.

**Choice of outcomes.** Multi-task learning is an approach to transfer domain knowledge contained in related outcomes and learn in paralleled using a shared representation [21]. The incentive for the approach is that the outcomes are reflections of different facets of a common latent representation. To better know the latent representation, we learn from different outcomes. When thinking of the patient state as the latent representation, we need to choose outcomes that are reflections of the patient state. In our study, antiplatelet, beta-blockers, and statins treatment are prescribed based on doctors' perception of the patient based on domain knowledge. MACE outcome is a direct result of the patient disease state. We thus include all four outcomes to better represent the patient. After assessing the classification performance, the patient representation better characterizes the patient state regarding MACE and antiplatelet treatment than the patient state regarding beta-blocker treatment.

*Table 4. Feature importance in each patient cluster*

| Cluster | 3 | 4 | 1 | 7 | 6 | 5 |
|---|---|---|---|---|---|---|
| Ethnic group (Han) | 0.046 | 0.070 | 0.048 | 0.050 | **0.127** | 0.088 |
| ACS Type (NSTEMI) | 0.029 | 0.015 | 0.030 | **0.039** | 0.016 | 0.008 |
| ACS Type (STEMI) | 0.002 | 0.002 | 0.003 | **0.003** | 0.002 | 0.002 |
| ACS Type (UA) | 0.085 | 0.134 | 0.059 | 0.045 | 0.138 | **0.208** |
| Current smoking | 0.046 | 0.052 | 0.091 | **0.095** | 0.032 | 0.030 |
| Current Hypertension | 0.032 | 0.103 | 0.036 | 0.033 | 0.127 | **0.146** |
| Current Diabetes | 0.032 | 0.033 | **0.053** | 0.034 | 0.034 | 0.028 |
| Current atrial fibrillation | 0.082 | 0.055 | 0.078 | **0.085** | 0.056 | 0.025 |
| Current Percutaneous coronary intervention | 0.064 | 0.071 | **0.101** | 0.086 | 0.018 | 0.032 |
| Elevated enzyme levels | 0.047 | 0.043 | 0.050 | 0.060 | **0.080** | 0.061 |
| Killip Class | **0.048** | 0.029 | 0.035 | 0.036 | 0.005 | 0.011 |
| Age (years) | **0.045** | 0.029 | 0.033 | 0.027 | 0.008 | 0.013 |
| Systolic blood pressure (mmHg) | **0.033** | 0.017 | 0.018 | 0.019 | 0.010 | 0.012 |
| White blood cell count (×10^9/L) | 0.029 | **0.031** | 0.020 | 0.020 | 0.019 | 0.029 |

**Pitfalls in interpretation of the results.** Two points need to be addressed regarding interpretation of the results. First, algorithmically meaningful clusters are not necessarily clinically meaningful clusters. Though the clustering result has implication for disease prognosis, whether it can inform clinical practice needs further clinical research. Second, risk factors cannot be directly translated to clinical intervention. As an example, though comorbidity of hypertension and high systolic blood pressure are risk factors for in-hospital MACE, the results are not sufficient to claim the intensity of hypertension treatment required for different clusters.

**Suggestions for future study.** Our suggestions for future study using similar approach include: (1) carefully select clinically meaningful outcomes to be used; (2) use fewer features and easily acquired ones would make the results more applicable; and (3) identify cluster-specific interventions considering treatment effectiveness would add extra clinical value to similar studies.

## Conclusions

In this study, we used multi-task neural network with attention as a modeling framework to support learning of patient state representations, cluster analysis of patients, and profiling of feature importance. Seven patient clusters were discovered, which have different characteristics and risk profile to in-hospital MACE. The results demonstrate cluster analysis using outcome-driven multi-task neural network as a promising approach for ACS patient classification and subtyping.

## Acknowledgements

This work was supported by the grants from National Science Foundation of China: 81530016 to Changsheng Ma, and from Beijing Municipal Commission of Science and Technology: D151100002215003 and D151100002215004 to Jianzeng Dong.

**Address for correspondence**

Eryu Xia, PhD, IBM Research – China, 28, Zhongguancun Software Park, Haidian District, Beijing, China. Email: xerbj@cn.ibm.com.